\documentclass[runningheads]{llncs}
\usepackage{graphicx}

\usepackage{mathtools}
\usepackage{amsfonts}
\usepackage{listings, xcolor}
\usepackage{caption}
\usepackage{subcaption}

\usepackage{float}
\colorlet{punct}{red!60!black}
\definecolor{background}{HTML}{EEEEEE}

\definecolor{eclipseKeywords}{rgb}{0.0, 0.0, 0.28}
\definecolor{delim}{rgb}{0.0,1.76,2.40}
\colorlet{numb}{magenta!60!blue}

\lstdefinelanguage{json}{
    basicstyle=\normalfont\footnotesize, 
    numbers=left,
    numberstyle=\footnotesize, 
    stepnumber=1,
    numbersep=8pt,
    showstringspaces=false,
    breaklines=true,
    frame=single,
    backgroundcolor=\color{white},
    stringstyle=\color{blue},
    string=[s]{"}{"},
    literate=
     *{0}{{{\color{red}0}}}{1}
      {1}{{{\color{red}1}}}{1}
      {2}{{{\color{red}2}}}{1}
      {3}{{{\color{red}3}}}{1}
      {4}{{{\color{red}4}}}{1}
      {5}{{{\color{red}5}}}{1}
      {6}{{{\color{red}6}}}{1}
      {7}{{{\color{red}7}}}{1}
      {8}{{{\color{red}8}}}{1}
      {9}{{{\color{red}9}}}{1}
      {:}{{{\color{red}{:}}}}{1}
      {*}{{{\color{delim}{*}}}}{1}
      {/}{{{\color{delim}{/}}}}{1}
      {,}{{{\color{red}{,}}}}{1}
      {...}{{{\color{red}{...}}}}{1}
      {\_}{{{\color{red}{\_}}}}{1}
      {\{}{{{\color{red}{\{}}}}{1}
      {\}}{{{\color{red}{\}}}}}{1}
      {[}{{{\color{red}{[}}}}{1}
      {(}{{{\color{red}{(}}}}{1}
      {)}{{{\color{red}{)}}}}{1}
      {]}{{{\color{red}{]}}}}{1},
}

\begin{document}
\title{Chaos and Logistic Map based Key Generation Technique for AES-driven IoT Security \thanks{Supported by RMIT RTS Program.}}
%
%
\author{Ziaur Rahman\inst{1} \and
Xun Yi\inst{1} \and
Ibrahim Khalil\inst{1} \and 
Mousumi Sumi\inst{2}}
\authorrunning{R. Ziaur et al.}
%
\institute{RMIT University, Melbourne VIC 3000, Australia \\
\email{\{rahman.ziaur, xun.yi, ibrahim.khalil\}@rmit.edu.au}\\
 \and
DSI Ltd., Dhaka 1206, Bangladesh\\
\email{mousumi.sumi@dsinnovators.com}}
\maketitle              
\begin{abstract}
Several efforts have been seen claiming the lightweight block ciphers as a necessarily suitable substitute in securing the Internet of Things.  Currently, it has been able to envisage as a pervasive frame of reference almost all across the privacy preserving of smart and sensor-oriented appliances. Different approaches are likely to be inefficient, bringing desired degree of security considering the easiness and surely the process of simplicity but security.  Strengthening the well-known symmetric key and block dependent algorithm using either chaos motivated logistic map or elliptic curve has shown a far-reaching potential to be a discretion in secure real-time communication. The popular feature of logistic maps, such as the un-foreseeability and randomness often expected to be used in dynamic key-propagation in sync with chaos and scheduling technique towards data integrity. As a bit alternation in keys, able to come up with oversize deviation, also would have consequence to leverage data confidentiality. Henceforth it may have proximity to time consumption, which may lead to a challenge to make sure instant data exchange between participating node entities. In consideration of delay latency required to both secure encryption and decryption, the proposed approach suggests a modification on the key-origination matrix along with S-box. It has plausibly been taken us to this point that the time required proportionate to the plain-text sent while the plain-text disproportionate to the probability happening a letter on the message made. In line with that the effort so far sought how apparent chaos escalates the desired key-initiation before message transmission. 

\keywords{Internet of Things \and AES Modification \and Key Generation Matrix \and Logistic map}
\end{abstract}
\section{Introduction}
Internet of Things (IoT) has its security issue which been able to pass its new born and infant stage and successfully entered into teenage stage \cite{s1}\cite{s6}\cite{s9}. Though it was born a several decades earlier than Advance Encryption Standard (AES), enabling these two different generation strangers sitting together with a view to confluencing each other has also been just out as well \cite{s2}\cite{s3}. Thus thinking that they are capable of co-aiding themselves especially through an aspect of data integrity and confidentiality has also not yet promisingly embraced with any noteworthy testament from any credible sources so forth known to us. In the face of extensive dubiety that often leads to a certain degree of perplexity among research community, Logistic Map incorporated AES has been emerged as a self-definable safe-guard to abolish the possible gap in the IoT security challenges \cite{s5}\cite{s8}. It has already been able to show its potentials in particular in this field by super-setting secure smart device authentication to ensure strong communication, decentralized data formulation or even automatic data purchasing. Thus, it is conceivably estimated that an emerging phenomenon of IoT utensils would be able to equip with the internet to ease the coming security aspect of intelligible and straightforward encryption and decryption. In IoT cluster head is device on the network contributes to reliable data transmission which in response accepts information before processing and encrypting them as well. We propose this algorithm can be applied on the cluster head of simple IoT network. Though heterogeneous complex network architecture has been rapidly evolving day by day, the flexibility deserves properly concerned as more often micro-shaped sensor devices are necessarily required in this area of security and privacy \cite{s13}\cite{s14}. 
Fig. 1 shows how Advanced Encryption Standard (AES) secures data generated from IoT sensors. However, though AES has proven security strength, its key generation technique can be broken if there is a desired computation system. Therefore, using conventional AES for critical and real-time data security brings challenges to data integrity.  From the motivation of improving the AES key generation technique, the proposed paper claims contributions as follows.

\begin{enumerate}
    \item It increases the Key generation complexity, thus reduces the chances of breaking the keys. Instead of incorporating conventional two-dimensional S-box, the proposed key generation technique uses the 3-Dimensional Key Generation Mechanism (3DKGM). To make it robust and usuable it is designed based on Chaos cryptography and Logistic Map.
    \item The proposed technique can preserve the integrity of sensitive  IoT data. The sensor should have the necessary computation capacity to adapt to the proposed system.
    \item The coding-based empirical and extensive evaluation justifies the proposed technique's security strength compared to similar approaches. 
\end{enumerate}

The rest of the paper is organized as follows; section II illustrates the background and related works. The following section includes the required technical preliminaries and the Applicability of Chaos-based key generation. Then section IV explains the proposed technique. Lastly, the paper consists of an empirical evaluation and initial cryptanalysis based on 3DKGM S-box. Finally, it concludes with the prospects of the paper. 

\begin{figure}[ht!]
\includegraphics[width=\textwidth]{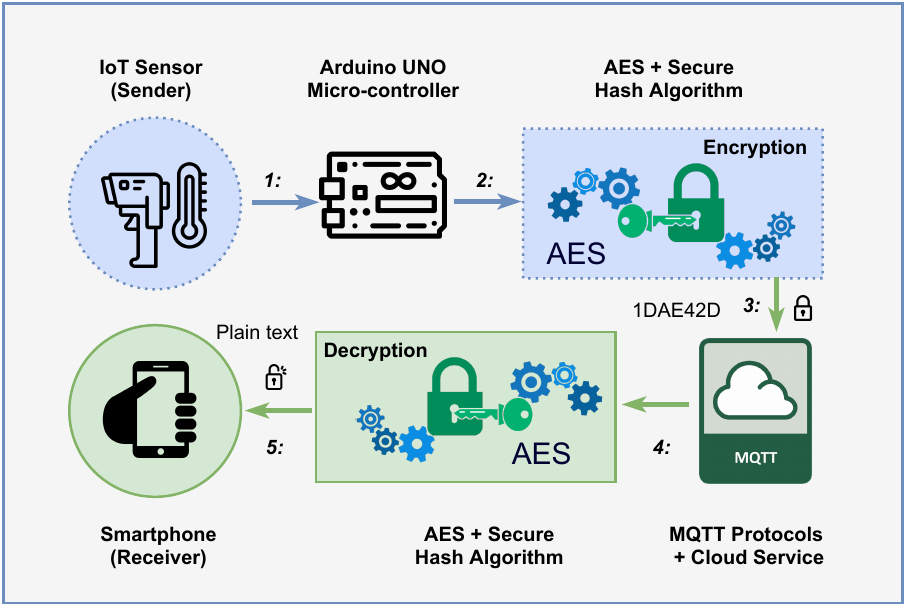}
\caption{How Advanced Encryption Standard (AES) is used to secure data generated from IoT sensors received by a smartphone.} \label{fig-new}
\end{figure}

\section{Background and Related Works}
The following portion of the paper briefly  explains the issues with the Symmetric Key algorithms focusing the Latency Challenges. Applying the symmetric-key algorithm in securing Internet of Things has been encountered enormous challenges [4][13], including potentials and opportunity. For the brevity only the succinct issues have been outlined hereby. 

\subsection{Latency Challenges}
A simple encryption algorithm can be represented as following:
\begin{equation}
    Y = E_zX
\end{equation}
Where $X$ is plaintext, $Y$ is cryptogram, $Z$ is a secret key, and $E_z$ is an encryption algorithm. To make any encryption algorithm workable on the devices associated, better it to simple and straightforward when we are in a time of introducing unfathomable advanced algorithms every day. In this area of secure communication certainly the principle concern should be required time to encrypt and decrypt the text messages.  Long time latency during this process may slow down system thus it decreases feasibility and usability of the system. It is known that each bit of plaintext conveys information. In addition to this it is even said that probability issue may help information achieving low latency. If it is said, `Today solar eclipse will occur', then it is an important matter, so the information of it is higher than the normal day. So, the bits of the plaintext is inversely proportional to occurring it. If the amount of information on the plaintext is $I$ and the probability of occurrence of that event is $P$, then 

\begin{equation}
I = f(x) =  \left\{
        \begin{array}{ll}
            0 & \quad if \quad P = 1 \\
            1 & \quad if\quad P = 0
        \end{array}
    \right.
\end{equation}

The important issue is presuming that, almost in every English message, letters are having a low probability of occurring such as $q, v, w, x, z$ lead to this point that they have low occurrence probability in accordance. If the probability of occurrences of letters $z$ and $r$ in and English message is $P_z$ and $P_r$ respectively, then it can be depicted as if-

\begin{equation}
P_r \leq P_z               
\end{equation}

It also can be said that the encryption of decryption time is much lower than which have higher probabilities. Not necessarily thinking about encryption and decryption time of letters has lower likelihood to occur. The concern of the proposed approach is those have a higher possibility to happen should have the reason for increasing the latency. Thus we would come up to introduce an algorithm that might be helping latency reduction for the letters with higher probability from a given message. We suggest that the letters have multiple occurrences may have the possibility in the text-message where maximum two letters will generate. In this situation, the popularly known $LZ78$ Algorithm \cite{s26} which will be described in the later section, has been considered to decrease the time of these types of letters.

\subsection{Related Works}
 There are several efforts have been made in the area application of the symmetric-key encryption techniques to make sure secure data communication considering lightweight \cite{s15}\cite{s32} and simplification approach \cite{s5}\cite{s10}. Baptista was one those \cite{s18} applied the concept of chaos \cite{s22} in the area of cryptography \cite{s23}\cite{s24}. Some other authors \cite{s27}\cite{s28} proposed if he could encrypt a message using low dimensional and chaotic logistic map though it was one-dimensional \cite{s25}\cite{s26}. Another work proposed to apply a single block of text message text, and it had low number of iterations and took longer time \cite{s16}\cite{s21}. For data encryption algorithm, chaotic map is one of the best ways of encryption for the high sensitivity of its initial condition \cite{s6}. In the continuous-time chaotic dynamic systems, poor synchronization and high noise problems may occur \cite{s7}. So far, logistic map is a 2-D chaotic map, but, more than logistic map, discretized 2-D chaotic maps are invented such as cat map, baker map and standard map \cite{s27} But the cat map and baker map have security issues, the other one \cite{s26}\cite{s28} the standard map are not thoroughly analyzed yet. LZ78 algorithm on the cryptography has been recently employed. It is used for source coding for lossless data compression [12][15]. Many more types of research have been done using LZ78 algorithm. But it is a rare case to use it as an application in the field of cryptography. Therefore, considering the chaotic properties, we have applied logistic map along with dynamic key generation matrix called 3-Dimensional Key Generation Matrix \cite{s19} to bring chaotic behavior inside. We are using it because we need faster and complex calculations to secure the algorithm and make it faster than before. The previous algorithm \cite{s19} secure with complex behavior, but we need more complex and faster procedures to cope with modern technology. So, we are dealing with chaos \cite{s20}.

\section{Technical Preliminaries and Applicability}
In response to the challenges outlined earlier, the next portion of the paper shows how several existing approaches incorporate AES for data integrity purpose. And the following sections explains the applicability of the Chaos based technique namely \textit{LZ78} algorithm. 

\subsection{Applying AES}
Now thinking about the prominent algorithm, one of the most common and ever robust encryption algorithms called Advanced Encryption Standard (AES) even though it was broken many times. Day by day, new and excellent modification approaches got invented and adapted to secure AES as well as to upgrade the accuracy, intruders, and their unauthorized access of information has become a typical phenomenon with the evolution of smart technology. Chaos based security [20] has been an essential concern in security research because of its randomness and unpredictability. User may not get any prior knowledge about initial condition means to discover desired key. A small variation will change the whole result; for example, modification 1 bit on plain text or key would bring a change the result nearly 50\%. Chaos-based cryptosystems are flexible for massive scale data such as audio and video in compare to the referred cryptosystems. Many authors have been trying to implement chaos in the existing cryptosystem \cite{s3} \cite{s20}. Chaos is dealt with the real numbers \cite{s4}, where other cryptographic methods deal with number of integers \cite{s5}. Thus, a chaos-based approach in the key-generation process could make the system much safer.

\subsection{Chaos Algorithms}
Chaos is fully exploited in the chaos-based cryptography. We can easily and safely transfer information using one dimensional logistic map[18[20][23]. The critical characteristics of chaos are to generate different intricate patterns and results in creating a large number of data by the mathematical model. These data can be used as secret keys [24]. The following procedures as shown in Fig. 2 are essential to stepping in the cryptographic algorithm- 

\begin{figure}[ht!]
\includegraphics[width=\textwidth]{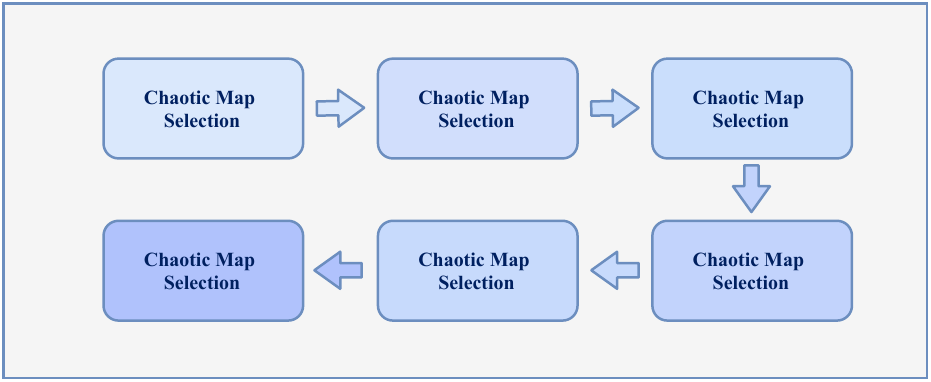}
\caption{A flow diagram of a generic Chaos Algorithm} \label{fig1}
\end{figure}

\subsection{LZ78 Algorithm}
\textit{LZ78} is a universal lossless data compression algorithm \cite{s26}. But we are going to use here to decrease the time of decryption. Because generally, the proposed algorithm is taking time in decrypting instead encrypting. Another problem is the decrypted result is not sometimes indexed correctly on the decrypted message. So, we are going to use $LZ78$ for proper indexing. It works from finding the probability of each letter of the text-message. A percentage will be generated to find out higher occurrence of a letter and lowest occurrence of letter. So, now if the message is `I Love to read book', then the encrypted message using $LZ78$ will be `I Love to read bo20k'. But, in this way the bits will be too significant than the actual message’s bits, which is not feasible. So, we modified this idea; rather than using $LZ78$ on letters, we are going to use it on each word of sentence. We are correctly indexing the word so that we could properly indexing the same words in a sentence. According to this concept, the above mentioned message will be the same `I Love to read book' because no other word is identical here. If the text-message is `Development process is a long process which helps to occur welfare development', then the message using $LZ78$ is `Development process is a long2 process which helps to occur welfare 1development'. We have organized the remaining of this paper as follows. In Section 2, a brief literature review of the chaotic standard map as discussed. The properties of analyzing it and a standard map for using in this paper is presented. The logic for choosing a logistic map and the key-generation process we proposed, and its working procedure as discussed in Section 3. In Section 4, algorithm selection for correcting position of the whole message. In Section 5, choosing logistic map and steps of key generating is discussed, and empirical evaluation and cryptanalysis is done in Section 6. Finally, in Section 7, some conclusions are drawn.

\section{Proposed Key Management Technique}
Handling the security in IoT environment is a challenging thing because of low power distribution, distributed nature, and no standardization. A concern of designing a security system is to assume that all the algorithms of cryptography are known to the attackers. Kerckhoffs's principle says that ‘secrecy of the key provides security.’ So, a key is more comfortable to protect from attacker than keeping secret the algorithm. And it is also a wise thing to keep confidential the key as it is a small piece of information. But trying to keep the keys secret is a challenging matter because a management system exists named ‘Key management.’ Key management is nothing but deals with the creation, modification, alternation, storage, practice, and replacement of keys; that means has access to internal access of keys. The critical management concerns not only the essentials of user level but also for the exchanges between end users. So, an algorithm is needed to interrupt the internal mechanism of key management procedure. Instead of using one key, nowadays more keys are generated, which are entirely dependent on each other. That may damage the security breach. So, this is the headache of today’s technology too. And the other one is brute force attack, which is bound in length of the key. To make the algorithm free from the brute force attack, [19] has experimented with it that may help to overcome it. This paper is concern about the increase the security breach where the key generated from the logistic map and keys are not wholly dependent on each other.

\subsection{Choosing Algorithm for Accurate Positioning}
Paper \cite{s1} works on AES, build a new key generation process using the 3DKGM matrix and S-box; both are three dimensional. All reduce the encrypting and decrypting time than existing AES because it tears off all the techniques that take time. But this paper uses RES method which is one of the most powerful and costing algorithms ever. Adding these operations to AES, the proposed algorithm gets additional computation to run and subsequently takes significant time. Although it has ensured about less timing, but this article aims at reducing the encrypting and decrypting the time than some other approaches. An encrypting algorithm is well-known when it concerns with both security and computational time (encrypting and decrypting time). To achieve both features, we have applied \textit{LZ78} algorithm properties and chaos theory. However, Advanced Encryption Standard (\textit{AES}) is one of the common block cipher algorithms, and the algorithm we propose deals with 35 bits, which is too long. After using \textit{s-box} and \textit{3DKGM} \cite{s19}, these huge number of bits turns into $744$ bits which is also long to handle with. But, we find the lowest and highest probability before applying LZ78 Algorithm on resulting plaintext. The word with the lowest probability is not created any big deal, but it will create big deal to the occurrence of highest probability. For example, if the message is `JONNY, your public key is equivalent to your address where you will receive cryptocurrency. So, keep your public key secret not to be interrupted on cryptocurrency.'. Now, the idea is that we have to find out lowest and highest probability of each word in a sentence and replace the higher probability with times of occurring in a sentence. As shown in the Table 1, the output for first sentence is $(0, JONNY,)$ $(0, your)$ $(0, public)$ $(0, key)$ $(0, is)$ $(0, equivalent)$ $(0, to)$ $(2, your)$ $(0, address)$ $(0, where)$ $(0, you)$ $(0, will)$ $(0, receive)$ $(0, cryptocurrency.)$ $(0, So,)$ $(0, keep)$ $(0, your)$ $(0, public)$ $(0, key)$ $(0, secret)$ $(0, to)$ $(0, not)$ $(7, to)$ $(0, be)$ $(0, interrupted)$ $(0, on)$ $(0, cryptocurrency.)$. 

\begin{table}
\caption{The LZ78 for the word `JONNY, your public key is equivalent to your address where you will receive cryptocurrency'}\label{tab1}
\begin{center}
\begin{tabular}{|l|c|l|}
\hline
\textbf{Output} &  \textbf{Index} & \textbf{String}\\
\hline
(0, JONNY,) &	1 &	JONNY, \\
\hline
(0, your) &	2 &	your \\
\hline
(0, public) & 3 & public \\
\hline
(0, key) & 4 & key \\
\hline
(0, is) &	5 &	is \\
\hline
(0, equivalent) &	6 &	Equivalent \\ \hline
(0, to) &	7 &	To \\ 
\hline
(2, your) &	8 &	Your \\ 
\hline
(0, address) & 	9 &	Address \\ 
\hline
(0, where) &	10 &	Where \\ 
\hline
(0, you) &	11 &	You \\ 
\hline
(0, will) &	12 &	Will \\ 
\hline
(0, receive) &	13 &	Receive \\ 
\hline
(0, cryptocurrency) &	14 &	cryptocurrency. \\ 
\hline

\end{tabular}
\end{center}
\end{table}

Now the plaintext that is processed removing $`0'$ is `JONNY; your public key is equivalent to 2your address you will receive cryptocurrency'. So, `keep your public key secret to not 7to be interrupted on cryptocurrency.' Now, it seems it taking a few bits more than real message, but it will be conducive for decrypting message.

\begin{figure}[ht!]
\centering
\includegraphics[width=6cm]{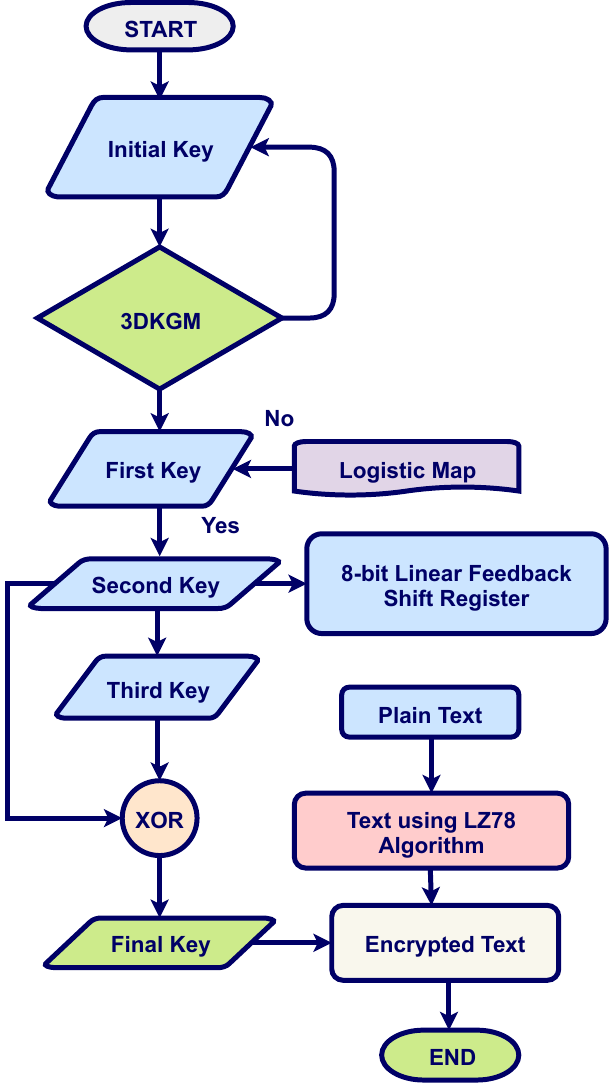}
\caption{Proposed key Generating Approach using Chaos} \label{fig2}
\end{figure}

\subsection{Choosing Logistic Map and Steps of Key Generating}
Different types of logistic maps have been proposed, and those are feasible. But to choose one, it must have three properties, such as- Mixing Property, Robust Chaos, Large Parameter \cite{s10}. By analyzing all the properties, we are going to use traditional logistic map. The equation is:
\begin{equation}
    X_{n+1} = rX_n(1-X_n)
\end{equation}

Where $X_n$ lies between zero and one, and the interval for $r$ is $[0,4]$. But, for the highly chaotic case, we are going to use $r=3.9999$. A key is the head of any block cipher encryption algorithm. In \cite{s19}, a fundamental generation matrix named \textit{3DKGM} (3-Dimensional Key Generation Matrix), which is the combination of Latin alphabets, integers, and Greek values used. Now, we want to extend the operation by using chaos. The reason is no one can get any prior knowledge about key. We are using here three keys. The first key gives birth of 1st key using \textit{3DKGM} \cite{s19}. The matrix looks like Fig. 4. The whole procedure of generating key is shown in Fig. 3.

\subsection{First Key Generation Process}
To use the matrix in the encryption algorithm is one of the tricky tasks. This type of job is building 3-Dimensional Key Generation Matrix. Let consider a secret key: `$POLY12@+\alpha \mu$'. At first, we have to declare the position of each byte. Then by using Fig. 4, we will get the first key, $k_1$. The procedure of getting this value as described in \cite{s19}. But it is needed to add that if any byte is missing from the table, three zero get replaced for the corresponding byte. The First Key generation process is depicted by Listing 1.1.

\begin{figure}[H]
\centering
\includegraphics[width=5cm]{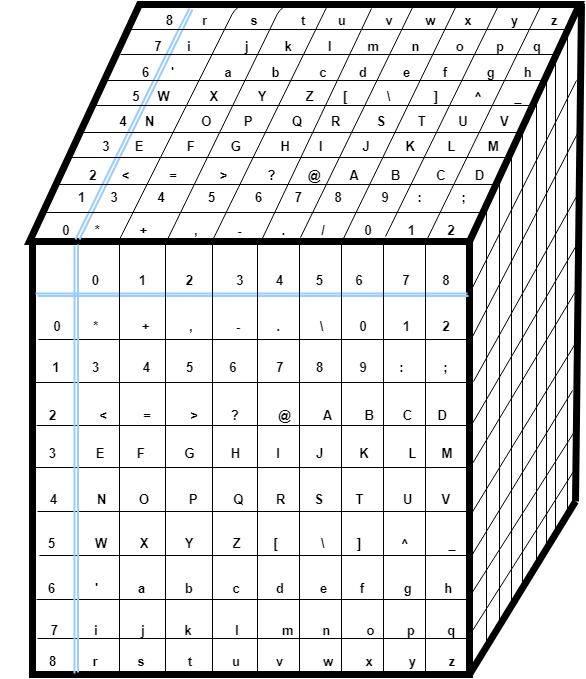}
\caption{3- Dimensional Key Generation Matrix (\textit{3DKGM}) \cite{s19}} \label{fig3}
\end{figure}

\begin{lstlisting}[language=json,caption=Pseudocode foro Algorithm 1: First Key Generation Process, firstnumber=1]
int cycle=0; /* algorithm starts */
for(int f=0;f<iniKey.length();f++) do
if(f<9) do cycle=f; end
if(f>=9) do cycle=0+cyy;
	cyy++; end
if(f>=18) do cycle=0+cyy1;
	   cyy1++; end
if(f>26) do cycle=0+cyy2;
	   cyy2++; end
if(f>=36) do cycle=0+cyy2;cycle=0+cyy3;
	   cyy3++; end
for(i=0;i<9;i++) do for(k=0;k<9;k++)
	  do for(k=0;k<9;k++)
	    do if(arr1[i][j][k]==iniKey.charAt(f))
		do hold1 = i;
		   hold2=j;end
	    end
    end 
end /* algorithm ends */
\end{lstlisting}

$k1= 42,41437C52G 07U08\_24d01j000000.$
Now, we have to calculate the initial condition from the logistic map.

\subsection{Find the Initial Condition from Logistic Map}
To find the initial condition $X_0$ from the logistic map, we have to choose first three blocks of first key. That is $42$,. And then convert it into corresponding binary number. For $4=00110100(B1)$, for $2=00110010$ \textit{(B2)} and for $,=00101100$ \textit{(B3)}. The mathematical representation of logistic map is

\begin{equation*}
\begin{aligned}
   X_i & = X_{i,j} \:\: where \:\: j=1,2\\
    X_{01} & = \frac{(B_1+B_2+B_3)}{2^{24}}\\
      & =(0\times2^0+ 0\times2^1+ 1\times2^2+ 1\times2^3+ 0\times2^4+ 1\times2^5+ 0\times2^6+ 0\times2^7)+\\
      & \:\:\:\:\:\: (0\times2^0+ 0\times2^1+1\times2^2+ 1\times2^3+ 0\times2^4+ 0\times2^5+ 1\times2^6+ 0\times2^7)+\\
      & \:\:\:\:\:\: (0\times2^0+ 0\times2^1+1\times2^2+ 1\times2^3+ 0\times2^4+ 0\times2^5+ 1\times2^6+ 0\times2^7)/2^{24}\\
      & =1.025 \times 10^{-05} \:\: (decimal)
  \end{aligned}
\end{equation*}

To calculate $X_{02}$, we need $4,5,6,7,8$ and $9$ blocks of the first key which is $41437C$.

\begin{equation*}
\begin{aligned}
    X_{02} & = \frac{(B_4+B_5+B_6+B_7+B_8+B_9)}{16\times 6}\\
      & =\frac{(00110100+00110001+00110100+00110011+00110111+01000011)}{96}\\
      & = 11 \\
      & = 3 \:\: (decimal)
  \end{aligned}
\end{equation*}
Therefore,
\begin{equation*}
\begin{aligned}
    X_0 & = (X_{01}+X_{02})\:\:mod\:\:1 \\
        & = 3 \:\: (decimal) \\
        & = =00000011 \:\: (binary)
  \end{aligned}
\end{equation*}

For the next cycle, $X_{01}$ will count from $B_{previous\_cycle} + 3$ to $B_{previous\_cycle}+5$. For the next cycle, $X_{02}$ will count from $B_{previous\_cycle}+6$ to $B_{previous\_cycle}+10$.

\subsection{Second and Third Key Generation Process}

Now, we will run several cycles to generate each byte of the second and third keys. From the above, we have found that
\begin{equation*}
    k_1= 42,41437C52G 07U08\_24d01j000000.
\end{equation*}

Now, we are taking the first block of first key which is $4$ (\textit{ASCII})$= 00110100$. From $X_0$, we have got the value is $3$, the binary value is $0000011$. So, $X_0 = 0000011$.Now, adding both of these binary values, we will get new key $K_2$:

\begin{equation*}
\frac{
    \begin{array}[b]{r}
       \:\:\:\:00000011 \\
      (+)\:\:\: 00110100 
    \end{array}
  }{
   K_2 = 00110111 
  }
\end{equation*}
To generate $K_3$, we need the help of $K_2$ and $8-$ bit linear feedback shift register. The functionality of $8-$ bit linear feedback shift register is to X-or the bits named $D_0$, $D_4$, $D_5$, $D_6$. After this operation, the answer is $0$. Now, we will do left shift operation on second key, that is, $0110111$. Then to put the bit on the right after $X-or$ operation: $01101110$. That is our desired third key, $K_3$ = $01101110$. Figure 5 (a) demonstrates the process. 

\begin{figure}
\centering
\includegraphics[width=\linewidth]{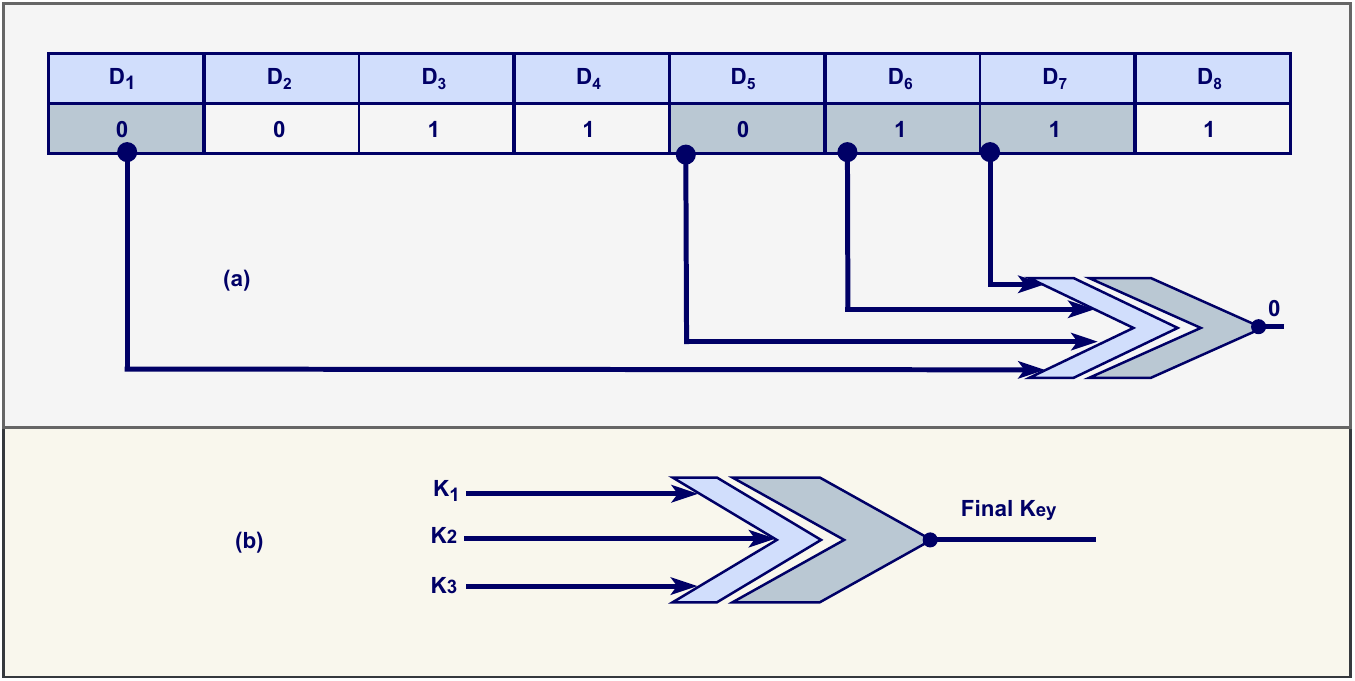}
\caption{(a) 8-bit linear feedback shift register (b) Generating final byte of the key} \label{fig4}
\end{figure}

\subsection{The Final Key}
After doing $X-or$ operation on the first key, second key, and third key, the first byte of final key will get generated. The final result is $01101110YYYYYYYY$.
So, this is the operation for one byte. Several bytes will get generated by using initial condition $X_0$ and the next byte of first key. Until the initial-key is covered, cycle will continue. After that plaintext and key will get into $x-or$ operation and follow the procedure from \cite{s19}. Therefore, each time, it generates bytes of key one by one. At last, It has to concatenate all the byte to build the final key which is going to the next step of encryption algorithm.Figure 5 (b) illustartes the Final key building process. 

\begin{figure}[ht!]
\centering
\includegraphics[width=10cm]{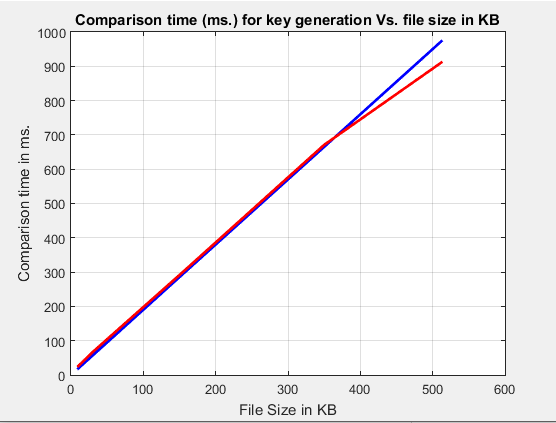}
\caption{Time-comparison of 2 Key generation techniques based on Table 2.}
\includegraphics[width=10cm]{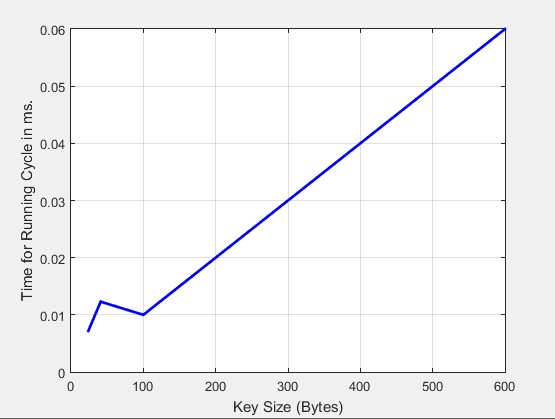}
\caption{Key Size vs Time for the Cycle Running based on Table 3.}
\end{figure}

\section{Initial Cryptanalysis and Empirical Evaluation}
The primary question of cryptography is security. The solution must be find theoretically and practically. The objectives of theoretical analysis are: increasing its randomness and computationally unpredictable. The objectives of practical analysis are to check up the above properties.
In \cite{s1}, we have used a key generation matrix named 3-Dimensional Key Generation Matrix. It is a static matrix which has a very similar working procedure with S-box. In \cite{s11}, for every $11$- round trail, it has $17$ active S-boxes, so, the differential trails.

\begin{equation*}
    DP \leq (2^{-4.678})^{17} \approx 2^{-79}
\end{equation*}

By analyzing this, we can also get a conclusion for the key in case of differential attacks and linear-attacks, as the trail of using 3DKGM depends on the length of initial key. As we use here the key is $POLY12@+\alpha\mu$, we have to use the matrix for this key is $10$ times, so the differential trails for $10$ rounds and And linear trails respectively are -

\begin{equation*}
    DP \leq (2^{-4.678})^{15} \approx 2^{-70}
\end{equation*}
\begin{equation*}
    LP \leq (2^{-4})^{15} \approx 2^{-60}
\end{equation*}

The paper \cite{s19} used $26$ rounds to gain ciphertext. Using \cite{s10}, we can answer its security. The standard security attacks are linear and differential. In \cite{s10}, $18$ round has minimum $27$ active bytes, so, for these $26$ rounds in \cite{s19}, it has $24$ active bytes. Therefore, $26$ rounds of differential trails.

\begin{equation*}
    DP \leq (2^{-4.678})^{24} \approx 2^{-112}
\end{equation*}

And for all $26$ rounds linear trails
\begin{equation*}
    LP \leq (2^{-4})^{24} \approx 2^{-96}
\end{equation*}

So, the linear and differential attacks are computationally infeasible. But it does not guarantee security from other attacks. There can be different type of security attacks too which needs further analysis. `Be on time' is a popular quote today. So, timing is a fact to encrypt and decrypt a message. As key is an essential and unavoidable part, so, it is needed to keep in mind to generate it at the lowest time. But, if it is taking a short time, then there may exist risk and can be broken by brute force process. Chaos is the best with the existing algorithm \cite{s19} to secure from all kinds of approaches. For experimental analysis, we use java for coding.

\subsection{Calculating Key Generation Time}
In this part, a comparison result of our proposed approach is shown with existing one. We have performed with different sizes of files and calculate the time of computation that shows the performance before and after adding chaos in the existing algorithm’s essential Generation process.

\begin{table}
\caption{Key Generation Time  vs. File Size  based on \cite{s19}}\label{tab3}
\begin{center}
\begin{tabular}{|c|c|c|}
\hline
\textbf{File size (KB)} &  \textbf{3DKGM key gen.(ms)} & \textbf{Proposed Chaos (ms) approach} \\
\hline
10 & 19 & 26 \\ \hline
30 &	57 &	67 \\ \hline
155 &	295 &	301 \\ \hline
350 &	665 &	671 \\ \hline
512 &	973 &	911 \\ \hline
\end{tabular}
\end{center}
\end{table}

\begin{table}
\caption{Time for Running Cycle for a Specific Key Size}\label{tab4}
\begin{center}
\begin{tabular}{|c|c|}
\hline
\textbf{Key Size (Bytes)} &  \textbf{Time for Running Cycle (ms)} \\
\hline
24 &	0.0072 \\ \hline
41 &	0.0123 \\ \hline
100 &	0.01 \\ \hline
255 &	0.02555 \\ \hline
300 &	0.003 \\ \hline
600 &	0.006 \\ \hline
\end{tabular}
\end{center}
\end{table}

Here from Table 2, we have taken file size of $10, 30, 155, 350$ and $512$.  And for $10, 30, 155, 350, 512$ \textit{kb} we get the corresponding time $26, 67, 301, 671, 911$. Fig. 5 demonstrates the further illustrations. So, the chaos-based approach decreases the computational time for long messages than the referenced one. So, it is giving two benefits, and it is giving strength and allows to encrypt the message within a little time. The respective Figure 6 represents the time comparison between key generation process. 


\subsection{Evaluating Time for Each Cycle}
In this part, in Table 3, we have shown a delay time corresponding to other cycles. If cycles are more, then it will take more time. But a reasonable time will be applicable. To note that, the period is entirely dependent on the length of key. Fig. 7 is showing a close view of delay of cycle for above mentioned. The x axis indicates number of bytes of key and y axis showing time in millisecond. For better visualization, the following Figure 6 is displayed.

\section{Future Perspective and Conclusion}
IoT sensors needs a security layer but has to be trusted enough for data integrity. Conventional AES has been proven to be insecure in several IoT cases. As known, AES security depends on S-box and key-scheduling which has significant impact on encryption and decryption. The proposed technique as demonstrated needs suitable chaotic map which lower the chances to break it. Based on chaos concept aligned with Logistic map, we have designed and demonstrated this novel key-scheduling process that has been designed to encrypt large volumes of data. Besides we have analyzed either it is secure against  different vulnerabilities.  The future scope includes justifying the further app;icability of the proposed scheme. As evaluated so far, the  proposed technique is safer for the IoT data integrity.
%
%
%

\begin{thebibliography}{8}
\bibitem{s1}
Song, T., Li, R., Mei, B., Yu, J., Xing, X. and Cheng, X.: A privacy preserving communication protocol for IoT applications in smart homes. IEEE Internet of Things Journal, 4(6), (2017). pp.1844-1852.

\bibitem{s2}
Moosavi, S.R., Gia, T.N., Nigussie, E., Rahmani, A.M., Virtanen, S., Tenhunen, H. and Isoaho, J.: End-to-end security scheme for mobility enabled healthcare Internet of Things. Future Generation Computer Systems, 64, (2016). pp.108-124.

\bibitem{s3}
Sehgal, A., Perelman, V., Kuryla, S. and Schonwalder, J.: Management of resource constrained devices in the internet of things. IEEE Communications Magazine, 50(12). (2012).

\bibitem{s4}
Mukhopadhyay, S.C. and Suryadevara, N.K.: Internet of things: Challenges and opportunities. In Internet of Things (pp. 1-17). (2014). Springer, Cham. 

\bibitem{s5}
Lee, I. and Lee, K. The Internet of Things (IoT): Applications, investments, and challenges for enterprises. Business Horizons, 58(4), ( 2015). pp.431-440. 

\bibitem{s6}
Yang, J., He, S., Lin, Y. and Lv, Z.: Multimedia cloud transmission and storage system based on internet of things. Multimedia Tools and Applications, 76(17), (2017). pp.17735-17750.

\bibitem{s7}
Farash, M.S., Turkanović, M., Kumari, S.: An efficient user authentication and key agreement scheme for heterogeneous wireless sensor network tailored for the Internet of Things environment. Ad Hoc Networks, 36, (2016). pp.152-176.

\bibitem{s8}
Saied, Y.B., Olivereau, A., Zeghlache, D. and Laurent, M.: Lightweight collaborative key establishment scheme for the Internet of Things. Computer Networks, 64, (2014). pp.273-295.

\bibitem{s9}
Jing, Q., Vasilakos, A.V., Wan, J., Lu, J. and Qiu, D.: Security of the Internet of Things: perspectives and challenges. Wireless Networks, 20(8), (2014). pp.2481-2501.

\bibitem{s10}
Yao, X., Chen, Z. and Tian, Y.: A lightweight attribute-based encryption scheme for the Internet of Things. Future Generation Computer Systems, 49, (2015). pp.104-112.

\bibitem{s11}
Turkanović, M., Brumen, B. and Hölbl, M., A novel user authentication and key agreement scheme for heterogeneous ad hoc wireless sensor networks, based on the Internet of Things notion. Ad Hoc Networks, 20, (2014). pp.96-112.

\bibitem{s12}
Sicari, S., Rizzardi, A., Grieco, L.A. and Coen-Porisini, A: Security, privacy and trust in Internet of Things: The road ahead. Computer networks, 76, (2015). pp.146-164. 

\bibitem{s13}
Ion, M., Zhang, J. and Schooler, E.M.: August. Toward content-centric privacy in ICN: Attribute-based encryption and routing. In Proceedings of the 3rd ACM SIGCOMM workshop on Information-centric networking (pp. 39-40). ACM. (2013).

\bibitem{s14}
Doukas, C. and Maglogiannis, I.: Bringing IoT and cloud computing towards pervasive healthcare. In Innovative Mobile and Internet Services in Ubiquitous Computing (IMIS), Sixth International Conference on (pp. 922-926). IEEE.(2012).

\bibitem{s15}
Beaulieu, R., Treatman-Clark, S., Shors, D., Weeks, B., Smith, J. and Wingers, L.: The SIMON and SPECK lightweight block ciphers. In Design Automation Conference (DAC), 2015 52nd ACM/EDAC/IEEE (pp. 1-6). IEEE. (2015).

\bibitem{s16}
Rahulamathavan, Y., Phan, R.C.W., Rajarajan, M., Misra, S. and Kondoz, A.: Privacy-preserving blockchain based IoT ecosystem using attribute-based encryption. In 2017 IEEE International Conference on Advanced Networks and Telecommunications Systems (ANTS) (pp. 1-6). IEEE. December. (2017).

\bibitem{s17}
Babar, S., Stango, A., Prasad, N., Sen, J. and Prasad, R., Proposed embedded security framework for internet of things (iot). In Wireless Communication, Vehicular Technology, Information Theory and Aerospace \& Electronic Systems Technology (Wireless VITAE), 2011 2nd International Conference on (pp. 1-5). IEEE. February. (2011).

\bibitem{s18}
Baptista, M.S.: Cryptography with chaos. Physics letters A, 240(1-2), (1998). pp.50-54. 

\bibitem{s19}
Rahaman, Z., Corraya, A.D., Sumi, M.A. and Bahar, A.N.: A Novel Structure of Advance Encryption Standard with 3-Dimensional Dynamic S-box and Key Generation Matrix. INTERNATIONAL JOURNAL OF ADVANCED COMPUTER SCIENCE AND APPLICATIONS, 8(2), (2017). pp.314-320. 

\bibitem{s20}
Kocarev, L.: Chaos-based cryptography: a brief overview. IEEE Circuits and Systems Magazine, 1(3), (2001). pp.6-21.

\bibitem{s21}
Guckenheimer, J. and Holmes, P.: Nonlinear oscillations, dynamical systems, and bifurcations of vector fields (Vol. 42). Springer Science \& Business Media. (2013).

\bibitem{s22}
Schneier, B.: Applied cryptography-protocols, algorithms, and source code in C. (1996).

\bibitem{s23}
Kotulski, Z., SZCZEPAŃSKI, J., Górski, K., Paszkiewicz, A. and Zugaj, A.: Application of discrete chaotic dynamical systems in cryptography—DCC method. International Journal of Bifurcation and Chaos, 9(06), (1999). pp.1121-1135.

\bibitem{s24}
Alvarez, G., Montoya, F., Romera, M. and Pastor, G.: Breaking parameter modulated chaotic secure communication system. Chaos, Solitons \& Fractals, 21(4), (2004). pp.783-787. 

\bibitem{s25}
Fridrich, J.: Symmetric ciphers based on two-dimensional chaotic maps. International Journal of Bifurcation and chaos, 8(06), (1998). pp.1259-1284. 

\bibitem{s26}
Ziv, J. and Lempel, A.: A universal algorithm for sequential data compression. IEEE Transactions on information theory, 23(3), (1977). pp.337-343. 

\bibitem{s27}
Lian, S., Sun, J. and Wang, Z.: A block cipher based on a suitable use of the chaotic standard map. Chaos, Solitons \& Fractals, 26(1), (2005). pp.117-129.

\bibitem{s28}
Kocarev, L. and Jakimoski, G.: Logistic map as a block encryption algorithm. Physics Letters A, 289(4-5), (2001). pp.199-206. 

\bibitem{s29}
Jakimoski, G. and Kocarev, L.: Differential and linear probabilities of a block-encryption cipher. IEEE Transactions on Circuits and Systems I: Fundamental Theory and Applications, 50(1), (2003). pp.121-123. 

\bibitem{s30}
Doukas, C. and Maglogiannis, I.: July. Bringing IoT and cloud computing towards pervasive healthcare. In Innovative Mobile and Internet Services in Ubiquitous Computing (IMIS), 2012 Sixth International Conference on (2012).  (pp. 922-926). IEEE. 

\bibitem{s31}
He, D. and Zeadally, S.: An analysis of rfid authentication schemes for internet of things in healthcare environment using elliptic curve cryptography. IEEE internet of things journal, 2(1), (2015). pp.72-83. 
\bibitem{s32}
Aziz, A. and Singh, K.: Lightweight Security Scheme for Internet of Things. Wireless Personal Communications, (2018). pp.1-17.

\bibitem{ref_article1}
Author, F.: Article title. Journal \textbf{2}(5), 99--110 (2016)

\bibitem{ref_lncs1}
Author, F., Author, S.: Title of a proceedings paper. In: Editor,
F., Editor, S. (eds.) CONFERENCE 2016, LNCS, vol. 9999, pp. 1--13.
Springer, Heidelberg (2016). \doi{10.10007/1234567890}

\bibitem{ref_book1}
Author, F., Author, S., Author, T.: Book title. 2nd edn. Publisher,
Location (1999)

\bibitem{ref_proc1}
Author, A.-B.: Contribution title. In: 9th International Proceedings
on Proceedings, pp. 1--2. Publisher, Location (2010)

\bibitem{ref_url1}
LNCS Homepage, \url{http://www.springer.com/lncs}. Last accessed 4
Oct 2017
\end{thebibliography}
%

\end{document}